# Margin setting with high-frequency data[1]

# John Cotter[2] and François Longin[3]

*Abstract*

Both in practice and in the academic literature, models for setting margin requirements in futures markets classically use daily closing price changes. However, as well documented by research on high-frequency data, financial markets have recently shown high intraday volatility, which could bring more risk than expected. This paper tries to answer two questions relevant for margin committees in practice: is it right to compute margin levels based on closing prices and ignoring intraday dynamics? Is it justified to implement intraday margin calls? The paper focuses on the impact of intraday dynamics of market prices on daily margin levels. Daily margin levels are obtained in two ways: first, by using daily price changes defined with different time-intervals (say from 3 pm to 3 pm on the following trading day instead of traditional closing times); second, by using 5-minute and 1-hour price changes and scaling the results to one day. Our empirical analysis uses the FTSE 100 futures contract traded on LIFFE.

Keywords: Clearinghouse, extreme value theory, futures markets, high-frequency data, intraday dynamics, margin requirements, risk management.

JEL Classification: G15

Version: 8.1 (July 27, 2006)

[1] The authors would like to thank Lianne Arnold, Kevin Dowd, David Hsieh (the Editor), John Knight, Andrew Lamb, Alan Marcus, and Hassan Tehranian for their helpful comments on this paper. The first author acknowledges financial support from a Smurfit School of Business research grant. The second author acknowledges financial support from the ESSEC research fund.

[2] Director of Centre for Financial Markets, Department of Banking and Finance, Smurfit School of Business, University College Dublin, Blackrock, Co. Dublin, Ireland, Tel.: +353-1-7168900. E-mail: john.cotter@ucd.ie.

[3] Professor of Finance, Department of Finance, ESSEC Graduate Business School, Avenue Bernard Hirsch B.P. 50105, 95021 Cergy-Pontoise Cedex, France. E-mail: contact@longin.fr. Web: www.longin.fr.

# 1. Introduction

The existence of margin requirements decreases the likelihood of customers' default, brokers' bankruptcy and systemic instability of futures markets. Margin requirements act as collateral that investors are required to pay to reduce default risk. [4] Margin committees face a dilemma however in determining the magnitude of the margin requirement imposed on futures traders. On the one hand, setting a high margin level reduces default risk. On the other hand, if the margin level is set too high, then the futures contracts will be less attractive for investors due to higher costs and decreased liquidity, and finally less profitable for the exchange itself. This quandary has forced margin committees to impose investor deposits that represent a practical compromise between meeting the objectives of adequate prudence and liquidity of the futures contracts.

Let us describe as an example the way margins are set on the London International Financial Futures and Options Exchange (LIFFE). For products traded on this exchange, margin requirements are set by the London Clearing House (LCH) (for further details see London Clearing House, 2002). The LCH risk committee is responsible for all decisions relating to margin requirements for LIFFE contracts. Margin committees generally involve experienced market participants who have widespread knowledge in dealing with margin setting and implementation, through their exposure to various market conditions and their ability to respond to changing environments (Brenner (1981)). The LCH risk committee is independent from the commercial function of the Clearinghouse. In order to measure and manage risk, the LCH uses the London Systematic Portfolio Analysis of Risk (SPAN) system, a specifically developed variation of the SPAN system originally introduced by the Chicago Mercantile Exchange (CME). The London SPAN system is a non-parametric risk-based model that provides output of margin requirements that are sufficient to cover potential default losses in all but the most extreme circumstances.[5] The inputs to the system are estimated margin requirements relying on price movements that are not expected to be exceeded over a day or couple of days. These estimated values are based on diverse criteria incorporating a focus on a contract's price history, its close-to-close price movements, its liquidity, its seasonality and forthcoming price sensitive events. Market volatility is specially a key factor to set margin levels. Most important however is the extent of the contract's price movements with a policy for a minimum margin requirement that covers three standard deviations of historic price volatility based on the higher of one-day or two-day price movements over the previous 60-day trading period. This is akin to using the Gaussian distribution where multiples of standard deviation cover certain price movements at various probability levels.[6]

Clearinghouses are also beginning to recognize the importance of intraday dynamics. For example, in 2002, the LCH has introduced an additional intraday margin requirement that is initiated if price movements on a contract challenge the prevailing margin requirement (London Clearing House, 2002). Specifically, an intraday margin requirement is initiated if a contract price changes by 65% of the margin requirement originally set for that contract.[7] In this case, the

---

[4] Futures exchanges also use capital requirements and price limits to protect against investor default.

[5] Alternative approaches in order to compute the margin requirement have been developed in the academic literature: Figlewski (1984), Gay *et al* (1986), Edwards and Neftci (1988), Warshawsky (1989), Hsieh (1993), Kofman (1993), Booth *et al* (1997), Longin (1999) and Cotter (2001) use different statistical distributions (Gaussian, historical or extreme value distribution) or processes (GARCH), Brennan (1986) proposes an economic model for broker cost minimization in which the margin is endogenously determined, and Craine (1992) and Day and Lewis (1999) model the distributions of the payoffs to futures traders and the potential losses to the futures clearinghouse in terms of the payoffs to barrier options.

[6] For instance, under the hypothesis of normality for price movements, two standard deviations would cover 97.72% of price movements, and three standard deviations 99.87%.

[7] The validity of the chosen cut-off point for imposing intraday margins cannot be taken for granted as it is arbitrarily chosen without any rational or justification.

Clearinghouse requires an additional margin payment for falling prices on a long position or for rising prices on a short position. The possible impact of intraday price movements is now clearly, and rightly so, of concern to risk management overseers for LIFFE contracts.

This paper tries to answer two questions relevant for margin committees in practice: is it right to compute margin levels based on closing prices and ignoring intraday dynamics? Is it justified to implement intraday margin calls? In order to answer these two questions this paper takes into account the intraday dynamics of futures market prices in computing margin requirements. All previous academic studies considered daily closing prices only, thus missing potentially important information. In our study we obtain daily margin levels in two ways: first, by using daily price changes defined with different time-intervals (say from 3 pm to 3 pm on the following trading instead of traditional closing times); second, by using 5-minute and 1-hour price changes and scaling the results to one day. The use of high frequency data may specially be beneficial in order to get more precise estimates of risk measures as shown by Merton (1980). The computation of risk management measures for futures at different frequencies has already been considered by Hsieh (1993).[8] Under the assumption of independence and identical distribution (iid), daily margin levels obtained over different time-intervals should be on average equal to and statistically different from daily margin levels obtained with closing prices. Identically, scaled intraday margin levels estimated with 5-minute and 1-hour price changes should be on average equal to daily margin levels obtained with closing prices. Any significant differences may then be accounted for by the lack of iid behavior. In such a case, it may be appropriate to set intraday margin levels in order to take into account specific intraday dynamics. In our paper, different statistical distributions are also used to model futures price changes: the Gaussian distribution, the extreme value distribution and the historical distribution. A GARCH process is also used to take into account the time-varying property of financial data. An application is given for the FTSE 100 futures contract traded on LIFFE.

---

[8] Hsieh (1993) computes long-term minimal capital requirements and daily minimum capital requirements while we look at short-term margins and daily margins. By focusing on the short term we adopt the position of the exchange concerned with its own risk while Hsieh (1993) takes the point of view of investors who may wish to hold their position for a long time and who are mainly concerned with the funding risk (in the case of a hedge).

The remainder of the paper is organized as follows. The statistical models used for the distribution of futures contract price changes and the scaling methods are presented in the next section. Section 3 provides a description of the FTSE 100 futures contract data used in the application and a detailed statistical analysis of the intraday dynamics of the market prices. Section 4 presents empirical results for margins by taking into account the intraday dynamics. Finally, a summary of the results and some implications for decision makers are given in the concluding section.

## 2. Statistical models and scaling methods

This section presents the different statistical models used to compute the margin level for a given probability. We do not necessarily select the best model but rather consider distributions that are used in practice by practitioners in charge of setting margins in derivative markets: the Gaussian and historical distributions (commonly used), the extreme value distribution (especially relevant for the problem of margin setting) and a GARCH type model (a conditional distribution). Our main goal is to study the impact of intra-day dynamics in margin levels and to show that such an impact is present whatever the distribution chosen. This section also presents the scaling method (where available) to obtain daily margin levels from intraday price changes.

**2.1 The Gaussian distribution**

The Gaussian distribution is considered because it is a standard tool in risk management. The unconditional Gaussian distribution of price changes requires the estimation of two parameters only, the mean, $\mu$, and the variance, $\sigma^2$. For a given probability $p$, the margin level corresponds to the quantile where one is examining what margin requirement is sufficient to exceed futures price changes over a time-period of length $T$ for the probability level $p$. Denoted by $ML(p, T)$, the margin level is computed as follows :

(1) $\quad ML(p,T) = \mu \cdot T + N^{-1}(p) \cdot \sigma \cdot \sqrt{T}$

where $N^{-1}$ is the inverse of the standardized Gaussian distribution.

As the expected price change can empirically be neglected over a short-time period (less than one day in our study), the scaling law relating the margin $ML(p, T)$ and the margin level for a basic time unit ($T=1$) follows the $\sqrt{T}$ rule:

(2) $\quad ML(p,T) = \sqrt{T} \cdot ML(p,1)$

**2.2 The extreme value distribution**

One question that we may ask about the nature of risk management is whether the clearinghouse should care more about ordinary market conditions or more about extraordinary market conditions. In other financial institutions such as banks two distinct approaches are used: value at risk models for ordinary market conditions and stress testing for extraordinary market conditions (see Longin (2000)). The clearinghouse must also address both sets of market conditions in margin setting so as to minimize the likelihood of investor default by examining a range of probabilities of price movements associated with common and uncommon events. For that reason the extreme value distribution is considered. It provides a precise model for the tail of the distribution of price changes.[9] Using the non-parametric estimation approach developed by Hill (1975), the margin level $ML(p, T)$ is computed as follows :

---

[9] See Embrechts *et al* (1997) for a presentation of extreme value theory.

$$(3) \quad ML(p,T) = r^{th} \cdot \left( \frac{n}{N \cdot p} \right)^{\frac{1}{\alpha}}$$

where $r^{th}$ is the tail threshold price change associated with the beginning of the sample of tail observations, $N$ the number of observations of price changes in the database, $n$ is the number of order statistics used to compute the tail parameter $\alpha$. The tail parameter measures the degree of tail thickness. It represents the number of bounded moments: moments lower than $\alpha$ are finite and moments equal to and greater than $\alpha$ are infinite. Extreme value studies applied to financial time-series (see Jansen and de Vries (1991) and Longin (1996) for example) have found tail parameter estimates between 2 and 4 suggesting that not all moments of the price changes are finite.

A result by Feller (1971) for the tail behavior under time-aggregation scales the results by using a $T^{1/\alpha}$ rule:

$$(4) \quad ML(p,T) = T^{\frac{1}{\alpha}} \cdot ML(p,1)$$

Importantly the tail parameter $\alpha$ remains invariant to the aggregation process and also has implications for empirical benefits in its actual estimation. Dacarogna *et al* (1995) have shown that high-frequency tail estimation has efficiency benefits due to their fractal behavior. In contrast, low frequency estimation suffers from negative sample size effects. Intuitively a large (high) frequency data set has more observable extremes that a small (low) frequency one over the same time interval thereby allowing for stronger inferences of these rare events. Furthermore for ease of computation, the scaling procedure does not require further estimation, but only involves parameters from the high-frequency analysis, shown to provide the most detailed information on futures price movements.

**2.3 Historical distribution**

The simplest way to calculate margins is as quantiles relying on the historical distribution of returns. This is also the method with the least model risk. The historical distribution provides margins representing a quantile using the full set of price changes ordered in ascending fashion:

(5) $\quad ML(p,T) = \min\left(r_n, \dfrac{n}{N} \geq 1-p\right)$

Note that there is no scaling law associated with the historical distribution and that we are limited to in-sample margin estimation.

**2.4 The GARCH process**

All statistical models presented above are based on unconditional distributions and cannot reflect current market conditions. As first noted by Hsieh (1991), modeling the conditional heteroskedasticity is a key point in the margin setting context. As market conditions may vary substantially over time, Hsieh suggests that the conditional density function may be used in a dynamic margin setting process. In order to take into account current market conditions a conditional process such as a GARCH process is used to address issues relating to the dynamic features of futures contracts volatility (see Cotter (2001)).[10] To model the time-varying behavior of price changes suggested by the previous analysis, we use the GARCH model developed by Bollerslev (1986) given by:

(6) $\quad \sigma_t = \alpha_o + \sum_{i=1}^{p} \alpha_i \varepsilon_{t-i}^2 + \sum_{j=1}^{q} \beta_j \sigma_{t-j}^2$

---

[10] See Hsieh (1993) for further applications of GARCH processes in modelling conditional density functions. Hsieh (1991) notes that the popularity of these models is due to their ability to capture the dependence structure of financial returns. Various potential explanations are given for this dependence structure resulting in non iid behaviour including deterministic chaos, non-stationarity and non-linear stochastic processes.

for $\alpha_0, \alpha_i, \beta_j \geq 0, \quad \alpha_i + \beta_j \leq 1.$

The unconditional level of volatility is related to $\alpha_0$, persistence in volatility of the innovations in $\varepsilon_{t-i}^2$ given by $\alpha_i$ and the persistence in past volatility $\sigma_{t-j}^2$ given by $\beta_j$.

A single lag GARCH (1, 1) model is applied here to the price series at the end of the sample during December 2000. Assuming the conditional distribution is Gaussian, results in scaling using the $\sqrt{T}$ rule outlined earlier.

## 3. Data analysis

### 3.1 Data

The empirical analysis is based on transaction prices for the FTSE 100 futures contract trading on the LIFFE exchange (data are obtained from *Liffedata*). This exchange has made a clear distinction, between contracts that are either linked to an underlying asset or developed formally on the basis of links to the recently developed European currency, the euro, and those that remain linked to factors outside the currency area. The FTSE 100 represents the most actively traded example of the latter asset type.

Data are available on the stock index contract for four specific delivery months per year, March, June, September and December. Prices are chosen from those contracts with delivery months on the basis of being the most actively traded using a volume crossover procedure. The empirical analysis is completed for sampling frequencies of 5 minutes, 1 hour and 1 day. The first interval is chosen so as to meet the objective of analyzing the highest frequency possible and capturing the most accurate risk estimates but also avoids microstructure effects such as bid-ask effects. For the daily frequency, the price changes are computed by taking different starting (and ending) times to define the day: the beginning of the "day" can start from 9 am (the opening of the trading day) to 5 pm (the closing of the trading day). Nine different time-series of daily price changes are then obtained. Log prices (or log prices to the nearest trade available) for each interval are first differenced to obtain each period's price change. The period of analysis is for the year 2000 involving 247 full trading days corresponding to an average life span of an exchange traded futures contract. The FTSE 100 futures daily interval encompasses 113 5-minute trading intervals and nine hourly trading intervals. A number of issues arise in the data capture process. First, all holidays are removed. This entails New Year's (2 days), Easter (2 days), May Day (1 day), spring holiday (1 day), summer holiday (1 day), and Christmas (2 days). In addition, trading took place over a half day during the days prior to the New Year and Christmas holidays and these full day periods are removed from the analysis.

**3.2 Basic statistics**

*Daily price changes defined with different time-intervals*

In addition to examining daily price changes using closing prices that are the norm in margin setting through the marking to market system, daily price changes can also be defined with different time-intervals. Basic statistics are reported in Table 1 and a time-series plot for two of these time-intervals, using opening prices and closing prices are presented in Figure 1. Whilst the mean price changes remain reasonably constant, other moments are more diverging suggesting the dynamics for different intervals vary. For instance, skewness goes from -0.09 to -0.47 and the kurtosis statistic goes from being platykurtic (-0.32) to leptokurtic (1.52). Also the dispersion of various quantiles is considerable. Again dependency varies according to the different time-intervals. Inferences for the squared price changes are similar although greater in magnitude. However it can be observed that both time-series have similar time-varying features evidencing volatility clustering with periods of high and low volatility but the diverging features are clearly demonstrated as suggested by the magnitude of realizations. For example, the maximum squared price change is equal to 9.79 for 4 pm and 34.13 for 9 am.

*Price changes defined with different frequencies*

Basic statistics are reported in Table 2 for price changes (Panel A) and for squared price changes (Panel B) defined with different frequencies. We find different statistical behavior according to the frequency of measurement with very strong dependency, excess kurtosis and clear lack of normality recorded at the highest intraday level (5-minute intervals). To begin, concentrating on the first four moments of the distribution, we find that kurtosis increases as the frequency increases. For price changes, the (excess) kurtosis is equal to 0.26 for a 1-day frequency, 1.54 for a 1-hour frequency and 254.50 for a 5-minute frequency. The high kurtosis (greater than the value equal to 0 implied by normality) gives rise to the fat-tailed property of futures price changes. It is also illustrated by the probability density function and QQ plots of the shapes of price changes for different frequencies given in Figure 2. The extent of fat-tails is strongest for 5-minute realizations supporting the summary statistics and this would impact tail quantiles (margins) for this frequency. Also, the magnitude of values for these realizations can be very large as indicated by the scale of the density plots. These features generally result in the formal rejection of a Gaussian distribution using the Kolmogorov-Smirnov test.[11] Deviations from normality are strongest at the highest frequency. The other moments emphasize the magnitude and scale of the realizations sampled at different frequencies. On average, price changes were negative during the year 2000 and unconditional volatility increases for interval size. Selected quantiles reinforce divergences in magnitude at different frequencies. Similar conclusions can be made for the proxy of volatility, the squared price changes, although the skewness and kurtosis are more pronounced. Moreover, autocorrelation changes dramatically according to frequency of estimation with much more dependency being recorded for 5-minute price changes. For instance, the Ljung-Box test statistic is 180.90 for 5-minute price changes, strongly rejecting the hypothesis of iid behavior, whereas in contrast this hypothesis is not rejected at daily frequency (at 5% confidence level). This is also verified for squared price changes.

---

[11] Whilst a formal rejection of normality for the full distribution of daily price is not recorded at common significance levels the tail behaviour in Figure 2 clearly indicates a fat-tailed property.

**3.3 Extreme value analysis**

Tail parameter estimates using different time-intervals to compute daily price changes are presented in Table 3 for the left tail (Panel A) and the right tail (Panel B). Following Huisman *et al* (2001) the point estimates are calculated using the weighted least squares technique that minimizes the small-sample bias (see the appendix for details of the estimation process). The point estimates range from 2.57 to 6.34 and the values are generally in line with previous findings (see Cotter (2001)). As the tail parameter is positive, the extreme value distribution is a Fréchet distribution that is obtained for a fat-tailed distribution of price changes.

The tail parameter is also estimated with higher frequency (Panel C). The tail parameter value seems to be stable under the temporal aggregation. It tends to increase as we move to higher frequency indicating a fatter tail recorded at intraday levels but this is not statistically significant. As expected, the precision is also much improved by using 5-minute and 1-hour price changes with lower standard errors. For example, for the left tail, the tail estimates with standard error in parentheses are: 2.77 (0.01) for 5-minute intervals, 2.83 (0.04) for 1-hour intervals and 3.11 (0.66) for daily intervals.

We also use the tail parameter estimates to test if the second and the fourth moment of the distribution are well defined. For classical confidence level (say 5%), we are unable to reject the hypothesis that the variance is infinite in any scenario, whereas we are able to reject the hypothesis that the kurtosis is infinite in many scenarios.

**3.4 Conditional estimation**

Time-varying behavior is described from fitting the GARCH model to both intraday price changes at 5-minute and 1-hour intervals and daily price changes from different time-intervals at the end of December 2000.[12] The GARCH estimates consistently indicate that the conditional distributions exhibit persistence, with past volatility impacting on current volatility as it is typical of GARCH modeling at daily intervals.[13] Furthermore the conditional distributions vary according to the time intervals analyzed that will give rise to different margin requirements.

# 4. Model-based margin requirements

This section presents empirical results for margin requirements obtained with daily price changes (4.1) and 5-minute and 1-hour price changes scaled to one day (4.2). In our analysis of margin requirements we are interested in two separate questions: should margin requirements be set with closing prices alone? Is there a justification for implementing intraday margin calls? We now turn to these questions.

**4.1 Margin requirement based on daily price changes**

Table 4 presents margin requirements obtained with daily price changes for a long position (Panel A) and for a short position (Panel B). Margin requirements are computed for a given probability. Four different values are considered: 95%, 99%, 99.6% and 99.8% corresponding to average waiting periods of 20, 100, 250 and 500 trading days.[14] Thinking of risk management for financial institutions, probabilities of 95% and 99% would be associated with ordinary adverse market events modeled by value at risk models, and probabilities of 99.6% and 99.8% with extraordinary adverse market events considered in stress testing programs. In the margin setting context, the probability reflects the degree of prudence of the exchange: the higher the probability, the higher the margin level, the less risky the futures contract for market participants, but the less attractive the contract for investors.

---

[12] Our application is given for illustrative purposes only. We could also have fit the GARCH model for full timeframe to obtain daily conditional margins throughout the year 2000.

[13] For instance the parameter estimates based on daily closing prices are: $\alpha_0 = 0.01$, $\alpha_1 = 0.01$, $\beta_1 = 0.96$.

Margin requirements are computed with the statistical models previously presented: three unconditional distributions (Gaussian, extreme value and historical) and a conditional process (the GARCH process). For the presentation of the results, the extreme value distribution will be the reference model as it presents many advantages (parametric distribution, limited model risk, limited event risk) and as the problem of margin setting is mainly concerned with extreme price changes. Beginning with the analysis of extreme value estimates, we first note that variation occurs in the estimates based on the different time-intervals to define daily price changes. For example, for a long position and a probability level of 95%, the estimated margin level ranges from 1.83% to 2.05% of the nominal position. For the most conservative level of 99.8%, it ranges from 2.77% to 5.32%, almost double. Also there does not seem to be a systematic pattern to these deviations. For instance, for a probability of 95%, the minimum is obtained with 2 pm prices and the maximum for closing prices, and for a probability of 99.8%, the minimum is obtained with 3 pm prices and the maximum for 10 am prices. The same remarks apply to a short position. These findings suggest that the daily price change distributions vary to some extent based on different time-intervals sampled suggesting separate tail behavior for each price series.

---

[14] The average waiting period for a given quantile (margin level) represents the time we have to wait on average to observe a price change greater than the margin level. As explained by Longin (2000), for high levels of risk, the concept of waiting period is more meaningful than a probability. For example, the difference in probability between 99.6% and 99.8% appears very small while translated in terms of waiting period, the associated daily margin events occur on average every year and every two years, which is easier to understand and relate to.

Turning to the Gaussian estimates, some key insights are obtained. First, the measures are almost identical for long and short positions due to the assumption of a symmetric distribution of futures price changes and an average price change close to zero over the period considered. In contrast, the extreme value distribution and the historical distribution take account of the possibility of non-symmetric features in line with the oft cited stylized facts of financial time series, and verified for the FTSE 100 futures contract of diverging upper and lower distribution shapes. However, in line with all the estimates, diverging margin estimates occur according to the time-intervals used to define price changes. For example, for a long position and a probability of 95%, the estimated margin varies from 1.94% using 4 pm prices to 2.21% using opening prices. Traditional comparisons of extreme value and normal risk estimates suggest the latter underestimates tail behavior due to its exponential tail decline that results in relatively thin-tailed features. These findings hold for the FTSE 100 contract for high probability levels of 99.6% and 99.8%. In contrast, for the relatively low probability level of 95%, this conclusion cannot be sustained and this is due to this confidence level representing a common rather than extreme threshold. For instance, the probability of this event occurring using daily data is once every 20 trading days representing a typical event rather than an extreme one, although it is the latter events that need to be guarded against to avoid investor default.

Then turning to the historical estimates, diverging margin requirements again occur according to the time-interval chosen with the largest (smallest) estimate on a long position at the 95% level happening at 1 pm (10 am). These estimates are based on using the historical price series gathered for the year 2000. The historical estimates are confined to in-sample inferences due to the limited number of price observations. This implies that margin setting using the historical distribution that tries to avoid investor default may not be able to model the events that actually cause the default, whereas in contrast, extreme value theory specifically models these tail values.

The margin requirements based on the unconditional distributions may be compared to the conditional estimates using the GARCH process. Again it is clear that estimation at different time-intervals necessitates diverging margins. For instance, the out-of-sample estimates measured at 11 am (3 pm) represent the largest (smallest) possible margin requirements for a long position. Comparing the extreme value and GARCH estimates provides information on the distinction between unconditional and conditional environments facing margin setters. Distinct patterns occur based on the volatility estimation for the last trading day of the sample (December 29, 2000).

Thus this analysis suggests that Clearinghouses should consider setting margin requirements based on different time-intervals so as to avoid ignoring intraday dynamics.

**4.2 Daily margin requirement estimated with high-frequency price changes**

Table 5 presents daily margin requirements obtained with 5-minute and 1-hour price changes for a long position (Panel A) and for a short position (Panel B). Margin levels are scaled to one day (see Section 2 for the presentation of the scaling method) and compared to the ones obtained directly from daily price changes (average of daily margin levels obtained with daily price changes defined on different time-interval as presented in Table 4). Different statistical models are used: three unconditional distributions (the Gaussian distribution, the extreme value distribution and the historical distribution) and a GARCH process. The historical estimates are sometimes not available (na) due to the lack of a scaling formula or to data unavailability for out-of-sample inferences.

The general conclusion that we can draw from the results presented in Table 5 is that daily margin levels estimated with higher frequency (5-minute and 1-hour price changes) are consistently higher than margin levels directly obtained from daily price changes. For the Gaussian and the extreme value distributions, daily margin levels estimated with 5-minute price changes are always higher than daily margin levels directly obtained with daily data. This is also true for margin levels estimated with 1-hour price changes (except once at the 95% probability level for the extreme value distribution). For example for Gaussian margins on a long position the average scaled high-frequency margin levels are approximately 50% higher than daily margin levels, which is significant from an economic point of view. A *t*-test also shows that this difference is significant from a statistical point of view. Similar findings hold for the extreme value distribution. The rationale for these results is as follows: the iid assumption of future price changes, which is used for scaling margin levels computed with high-frequency data is not verified in practice.

The Clearinghouse must address the implication of these findings. One way is to introduce intraday margins that require additional payments from futures traders based on intraday price movements. As we have seen, intraday price movements are not correctly reflected in daily margins using scaling laws and this would encourage the Clearinghouse to have an additional payments system for traders to protect against these (extreme) price movements.

## 5. Summary and economic implications

This paper takes into account the intraday dynamics of futures prices changes in margin setting. It then includes lost information that is unavailable with the traditional approach of using closing prices in a marking to market system. The intraday futures price movements are relied on in two ways. First, daily price changes defined with different time-intervals are used to compute daily margin levels, and second high-frequency 5-minute and 1-hour price changes are used to compute intraday margin levels that are then scaled to give daily margin levels.

This paper finds that intraday dynamics should be a key component in margin setting. Daily price movements measured at different intervals can have a very tenuous relationship suggesting that the common procedure of using only close of day prices neglects the dynamics that investors actually face in trading futures. Daily margin levels estimated with high-frequency data (5-minute and 1-hour price changes) are consistently higher than daily margin levels directly obtained from daily price changes. Under the basic assumption of an iid process for price changes, which is used for the scaling law, margin levels based on high-frequency data should be more precisely estimated (it is the case) but on average not different from margin levels directly obtained from daily price changes.

The two economic issues pointed out in the introduction of this paper were about the use of closing prices to set daily margin levels and the justification of imposing intraday margins. Let us consider first the issue of setting daily margins. A margin level computed with closing prices may be substantially different from a margin level computed with another time-interval. The same result is obtained with the scaled daily margins from high-frequency price changes, which are substantially higher than the average margin level based on daily price changes. When deciding about the daily margin level and taking into account the intraday dynamics of price movements, the margin committee may consider margin levels computed with different time-intervals. A conservative approach would lead to considering the highest margin level over all time-intervals. The margin committee may also set daily margin level based on scaled margin levels from high-frequency price changes. Note that the empirical study carried out in this paper shows that both approaches (highest daily margin level based on price changes computed with different time-intervals and scaled daily margin levels based on high-frequency price changes) would lead to very similar values for daily margin levels. Indeed both approaches take into account the intraday dynamics in different but related ways. Let us consider now the issue of intraday margins. This paper shows that if the margin committee set daily margin levels by considering closing prices alone, it would to underestimate the margin level for a given level of risk. Then it makes sense to add intraday margins in order to take into account the extra risk due to the intraday dynamics. From a decision making point of view, the overall conclusion of this paper is that by not accounting for intraday dynamics the Clearinghouse may set inadequate margins resulting in unexpected high levels of default risk.

# Appendix

## Estimation of the tail parameter

This appendix describes the estimation procedure for the tail parameter of the extreme value distribution.

We use the method developed by Hill (1975) to estimate the tail parameter and also to determine the distribution quantiles (margin levels). The Hill estimator is widely used in empirical studies as it performs well for most time-series (Hall and Welsh (1984)) and is more efficient than other estimators based on order statistics (Kearns and Pagan (1997)). It is used in our scaling procedure for the extreme value method (Dacarogna *et al*, 1995)The Hill estimator corresponds to the maximum likelihood estimator of the inverse of the tail parameter $1/\alpha$:

(A1) $\quad \dfrac{1}{\alpha} = \dfrac{1}{n}\sum_{i=1}^{n}\left(\ln r_{N+1-i} - \ln r_{N-n}\right)$

De Haan *et al* (1994) shows that this tail estimator is asymptotically normal.

The issue in the estimation procedure is the choice of the optimal number of tail observations (*n*) to include in the estimator (see Danielson *et al* (2001) for a discussion). The dilemma faced is that there is a trade-off between the bias and variance of the estimator with the bias decreasing and the variance increasing with the number of tail observations used. In order to choose the optimal number of tail observations, we apply the regression method introduced by Huisman *et al* (2001):

(A2) $\quad \dfrac{1}{\alpha}(n) = \beta_0 + \beta_1 n + \varepsilon(n), \quad n = 1,\ldots,\eta$

For a weighted least squares regression of Hill estimates against associated numbers of tail estimates that minimizes heteroskedasticity in the regression's error term. Huisman et al. (2001) find that the estimator works well from simulation of small samples (similar in size to that analyzed here for daily intervals).

# References:

Bollerslev, T., Chou, R.Y. Kroner, K.F., 1992. ARCH Modeling in Finance: A Review of the Theory and Empirical Evidence, Journal of Econometrics, 52, 5-59.

Bollerslev T., Cai, J., Song, F.M., 2000. Intraday Periodicity, Long Memory Volatility, and Macroeconomic Announcement Effects in the US Treasury Bond Market, Journal of Empirical Finance, 7, 37-55.

Booth G.G., Broussard, J.P., Martikainen, T, Puttonen, V., 1997. Prudent Margin Levels in the Finnish Stock Index Market, Management Science, 43, 1177-1188.

Brennan, M.J., 1986. A Theory of Price Limits in Futures Markets, Journal of Financial Economics, 16, 213-233.

Brenner T.W., 1981. Margin Authority: No Reason for a Change, Journal of Futures Markets, 1, 487–490.

Cotter J., 2001. Margin Exceedances for European Stock Index Futures using Extreme Value Theory, Journal of Banking and Finance, 25, 1475-1502.

Cotter J., 2004. Minimum Capital Requirement Calculations for UK Futures, Journal of Futures Markets, 24, 193-220.

Craine R., 1992. Are Futures Margins Adequate ?, Working Paper, University of California – Berkley.

Dacarogna M.M., Pictet, O.V., Muller, U. A., de Vries, C.G., 1995. Extremal Forex Returns in Extremely Large Data Sets, Mimeo, Tinbergen Institute.

Danielson J., de Haan, L., Peng, L., de Vries, C.J., 2001. Using a Bootstrap Method to Choose the Sample Fraction in Tail Index Estimation, Journal of Multivariate Analysis, 76, 226-248.

Day T.E. Lewis, C.M., 1999. Margin Adequacy and Standards: An Analysis of the Crude Oil Futures Markets, Working Paper, Owen Graduate School of Management, Vanderbilt University.

Edwards F.R., Neftci, S.N., 1988. Extreme Price Movements and Margin Levels in Futures Markets, Journal of Futures Markets, 8, 639-655.

Feller W., 1971. An Introduction to Probability Theory and its Applications, John Wiley, New York.

Figlewski S., 1984. Margins and Market Integrity: Margin Setting for Stock Index Futures and Options, Journal of Futures Markets, 4, 385–416.

Gay G.D., Hunter, W.C., Kolb, R.W., 1986. A Comparative Analysis of Futures Contract Margins, Journal of Futures Markets, 6, 307–324.

De Haan L.S., Jansen, D.W., Koedijk, K., de Vries, C.G., 1994. Safety First Portfolio Selection, Extreme Value Theory and Long Run Asset Risks, in Galambos, C. (Ed.), Proceedings from a Conference on Extreme Value Theory and Applications. Kluwer Academic Publishing, Dordrecht, 471–487.

Hall P., Welsh, A., 1984. Best Attainable Rates of Convergence for Estimates of Parameters of Regular Variation, Annals of Statistics, 12, 1072-1084.

Hsieh D.A., 1991. Chaos and Nonlinear Dynamics: Application to Financial Markets, Journal of Finance, 46, 1839-1877.

Hsieh D.A., 1993. Implications of Nonlinear Dynamics for Financial Risk Management, Journal of Financial and Quantitative Analysis, 28, 41-64.

Huisman R., Koedijk, K., Kool, C.J.M., Palm, F., 2001. Tail Index Estimates in Small Samples, Journal of Business and Economic Statistics, 19, 208-215.

Jansen D.W., De Vries C.G., 1991. On the Frequency of Large Stock Returns: Putting Booms and Busts into Perspectives, Review of Economics and Statistics, 73, 18-24.

Kearns P., Pagan, A., 1997. Estimating the Density Tail Index for Financial Time Series, Review of Economics and Statistics, 79, 171-175.

Kofman P., 1993. Optimizing Futures Margins with Distribution Tails, Advances in Futures and Options Research, 6, 263–278.

London Clearing House, 2002, Market Protection, The role of LCH: regulatory framework, structure and governance, legal and contractual obligations, risk management, default rules, financial backing. London: LCH.

Longin F.M., 1996. The Asymptotic Distribution of Extreme Stock Market Returns, Journal of Business, 63, 383-408.

Longin F.M., 1999. Optimal Margin Levels in Futures Markets: Extreme Price Movements, Journal of Futures Markets, 19, 127-152.

Longin F.M., 2000. From Value at Risk to Stress Testing: The Extreme Value Approach, Journal of Banking and Finance, 24, 1097-1130.

Merton R.C., 1980. On Estimating the Expected Return on the Market, Journal of Financial Economics, 8, 323-361.

Warshawsky M.J., 1989. The Adequacy and Consistency of Margin Requirements: The Cash, Futures and Options Segments of the Equity Markets, Review of Futures Markets, 8, 420-437.

**Figure 1. FTSE 100 futures contract daily price changes and squared price changes defined with opening and closing prices.**

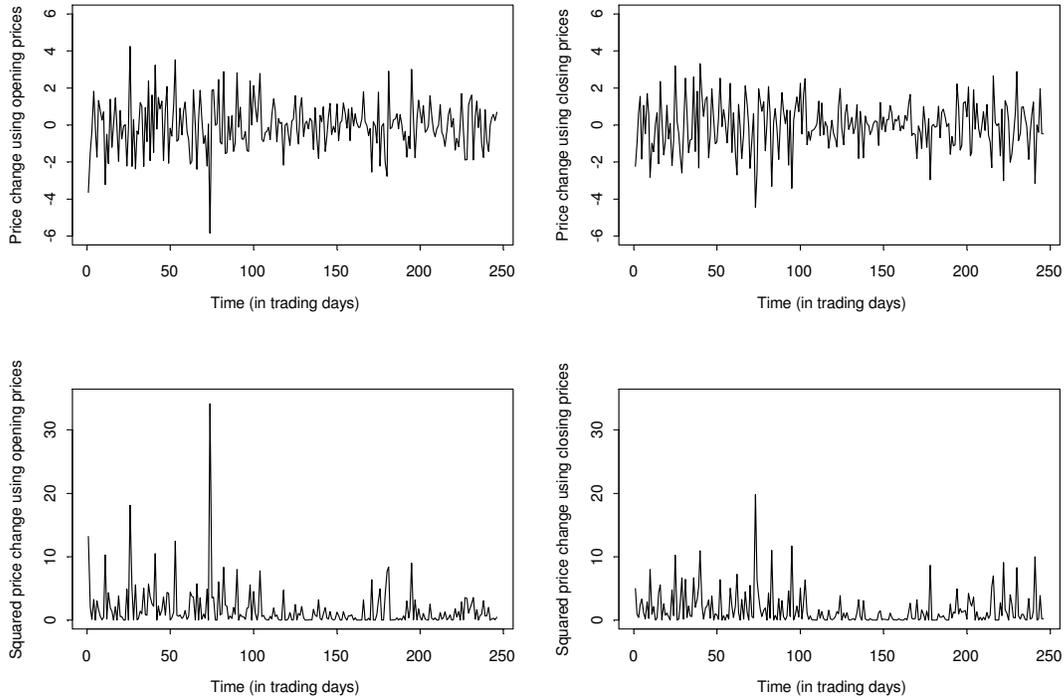

*Note:* these figures represent the history of the price change and squared price change of the FTSE 100 futures contract for the year 2000. Daily price changes are computed in two ways: from 9 am to 9 am on the following day (opening prices) and from 5 pm to 5 pm (closing prices).

**Figure 2. Probability density function and QQ plot for the FTSE 100 futures contract price changes defined with different frequencies.**

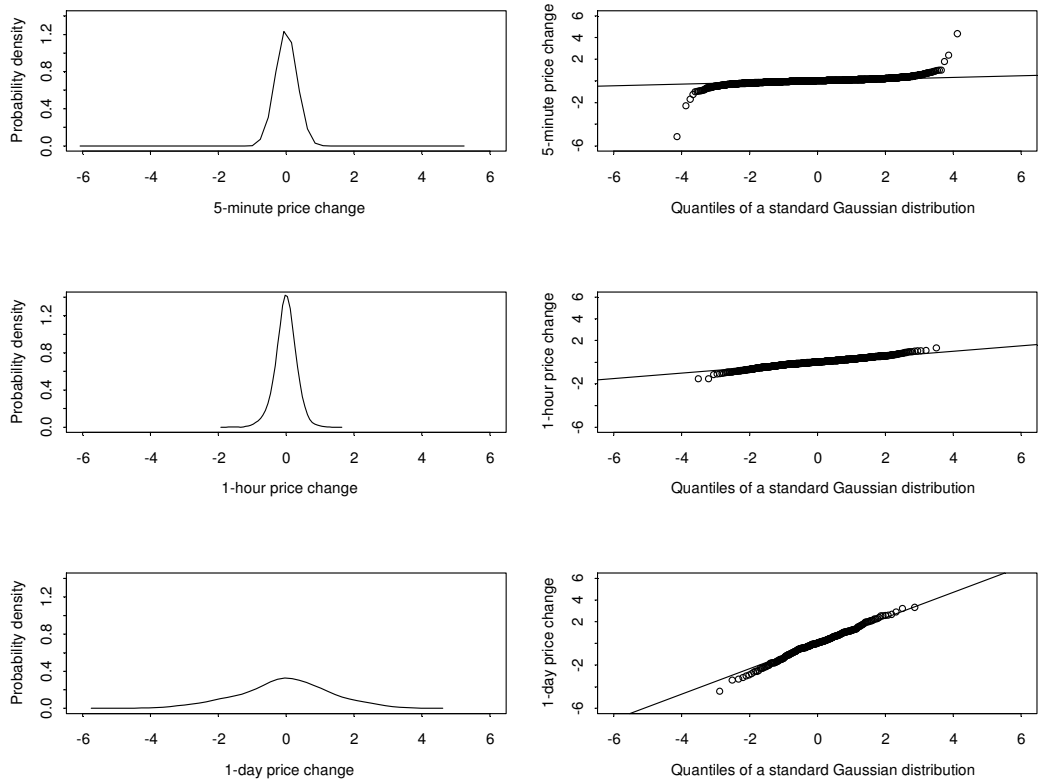

*Note:* these figures represent the probability density function and the QQ plots for price changes in the FTSE 100 futures contract for the year 2000. Three different frequencies are used to compute the price changes: 5 minutes, 1 hour and 1 day.

**Table 1. Basic statistics for the FTSE 100 futures contract daily price changes defined with different time-intervals.**

**Panel A. Price changes**

|  | Open | 10 am | 11 am | 12 pm | 1 pm | 2 pm | 3 pm | 4 pm | Close |
|---|---|---|---|---|---|---|---|---|---|
| Mean | -0.04 | -0.04 | -0.03 | -0.03 | -0.03 | -0.03 | -0.04 | -0.03 | -0.03 |
| Standard deviation | 1.32 | 1.23 | 1.20 | 1.23 | 1.18 | 1.29 | 1.22 | 1.16 | 1.30 |
| Skewness | -0.13 | -0.10 | -0.30 | -0.47 | -0.32 | -0.13 | -0.14 | -0.09 | -0.15 |
| Kurtosis | 1.52 | 1.13 | 0.88 | 1.39 | 0.16 | 0.14 | -0.05 | -0.32 | 0.26 |
| Kolmogorov-Smirnov | 0.05 | 0.04 | 0.05 | 0.06 | 0.04 | 0.03 | 0.04 | 0.03 | 0.04 |
| test of normality | (0.10) | (0.48) | (0.11) | (0.11) | (0.46) | (0.62) | (0.57) | (0.71) | (0.31) |
| Ljung-Box | 26.29 | 26.29 | 34.98 | 32.83 | 34.25 | 29.83 | 41.28 | 36.47 | 31.68 |
| test of white noise | (0.16) | (0.16) | (0.02) | (0.04) | (0.02) | (0.07) | (0.00) | (0.01) | (0.05) |
| Minimum | -5.84 | -4.92 | -4.74 | -5.73 | -4.48 | -4.54 | -3.60 | -3.13 | -4.38 |
| $1^{st}$ quartile | -0.79 | -0.86 | -0.78 | -0.76 | -0.80 | -0.79 | -0.79 | -0.80 | -0.77 |
| $2^{nd}$ quartile | -0.04 | -0.01 | 0.02 | -0.01 | 0.03 | -0.02 | -0.04 | 0.02 | 0.00 |
| $3^{rd}$ quartile | 0.78 | 0.74 | 0.73 | 0.81 | 0.80 | 0.86 | 0.78 | 0.76 | 0.76 |
| Maximum | 4.26 | 4.06 | 3.59 | 3.09 | 2.59 | 3.20 | 3.02 | 2.48 | 3.20 |

**Panel B. Squared price changes**

|  | Open | 10 am | 11 am | 12 pm | 1 pm | 2 pm | 3 pm | 4 pm | Close |
|---|---|---|---|---|---|---|---|---|---|
| Mean | 1.73 | 1.51 | 1.44 | 1.51 | 1.40 | 1.65 | 1.48 | 1.35 | 1.70 |
| Standard deviation | 3.24 | 2.66 | 2.44 | 2.79 | 2.06 | 2.42 | 2.06 | 1.74 | 2.55 |
| Skewness | 5.38 | 4.49 | 4.27 | 6.58 | 4.03 | 3.15 | 2.25 | 1.75 | 2.69 |
| Kurtosis | 43.77 | 27.90 | 26.24 | 65.77 | 27.84 | 16.08 | 5.94 | 2.90 | 10.38 |
| Kolmogorov-Smirnov | 0.30 | 0.29 | 0.28 | 0.29 | 0.25 | 0.25 | 0.24 | 0.22 | 0.25 |
| test of normality | (0.00) | (0.00) | (0.00) | (0.00) | (0.00) | (0.00) | (0.00) | (0.00) | (0.00) |
| Ljung-Box | 44.01 | 34.91 | 21.78 | 21.00 | 44.36 | 31.70 | 55.00 | 40.73 | 29.85 |
| test of white noise | (0.00) | (0.02) | (0.35) | (0.40) | (0.00) | (0.05) | (0.00) | (0.00) | (0.07) |
| Minimum | 0.00 | 0.00 | 0.00 | 0.00 | 0.00 | 0.00 | 0.00 | 0.00 | 0.00 |
| $1^{st}$ quartile | 0.09 | 0.13 | 0.13 | 0.14 | 0.13 | 0.18 | 0.16 | 0.12 | 0.14 |
| $2^{nd}$ quartile | 0.63 | 0.62 | 0.58 | 0.60 | 0.64 | 0.72 | 0.63 | 0.60 | 0.58 |
| $3^{rd}$ quartile | 2.05 | 1.70 | 1.74 | 1.94 | 1.85 | 1.86 | 1.95 | 1.76 | 2.21 |
| Maximum | 34.13 | 24.24 | 22.46 | 32.87 | 20.06 | 20.60 | 12.93 | 9.79 | 19.17 |

*Note:* this table gives the basic statistics and empirical quantiles for price changes (Panel A) and squared price changes (Panel B) defined with different time-intervals. It also presents the results of the Kolmogorov-Smirnov test for normality and the Ljung-Box Q-statistic for white noise with the *p*-value below in parentheses. To define the price change, the starting time, which is equal to the ending time on the following day, varies from 9 am (opening of the market) to 5 pm (closing of the market). Data are price changes of the FTSE 100 futures contract over the year 2000.

**Table 2. Basic statistics for the FTSE 100 futures contract daily price changes defined with different frequencies.**

**Panel A. Price changes**

|  | Frequency of price changes | | |
|---|---|---|---|
|  | 5-minutes | 1-hour | 1-day |
| Mean | 0.00 | -0.02 | -0.03 |
| Standard deviation | 0.11 | 0.30 | 1.30 |
| Skewness | -1.44 | -0.28 | -0.15 |
| Kurtosis | 254.5 | 1.54 | 0.26 |
| Kolmogorov-Smirnov | 0.08 | 0.05 | 0.04 |
| test of normality | (0.00) | (0.00) | (0.31) |
| Ljung-Box | 180.90 | 73.75 | 31.68 |
| test of white noise | (0.00) | (0.00) | (0.05) |
| Minimum | -5.17 | -1.57 | -4.38 |
| $1^{st}$ quartile | -0.05 | -0.18 | -0.77 |
| $2^{nd}$ quartile | 0.00 | -0.00 | -0.03 |
| $3^{rd}$ quartile | 0.05 | 0.16 | 0.76 |
| Maximum | 4.34 | 1.29 | 3.20 |

**Panel B. Squared price changes**

|  | Frequency of price changes | | |
|---|---|---|---|
|  | 5-minutes | 1-hour | 1-day |
| Mean | 0.01 | 0.09 | 1.70 |
| Standard deviation | 0.21 | 0.17 | 2.55 |
| Skewness | 107.99 | 5.24 | 2.69 |
| Kurtosis | 12 815.78 | 46.5 | 10.38 |
| Kolmogorov-Smirnov | 0.47 | 0.29 | 0.25 |
| test of normality | (0.00) | (0.00) | (0.00) |
| Ljung-Box | 6 351.26 | 107.51 | 29.85 |
| test of white noise | (0.00) | (0.00) | (0.07) |
| Minimum | 0.00 | 0.00 | 0.00 |
| $1^{st}$ quartile | 0.00 | 0.01 | 0.14 |
| $2^{nd}$ quartile | 0.00 | 0.03 | 0.65 |
| $3^{rd}$ quartile | 0.01 | 0.09 | 2.21 |
| Maximum | 26.73 | 2.46 | 19.17 |

*Note:* this table gives the basic statistics and empirical quantiles for price changes (Panel A) and squared price changes (Panel B) defined with different frequencies. Three different frequencies are used to compute the price changes: 5 minutes, 1 hour and 1 day. The table also presents the results of the Kolmogorov-Smirnov test for normality and the Ljung-Box Q-statistic for white noise with the *p*-value below in parentheses. Data are price changes of the FTSE 100 futures contract over the year 2000.

**Table 3. Tail parameter estimates and test of the existence of moments for the FTSE 100 futures contract price changes.**

**Panel A. Daily future price changes - Left tail**

|  | Open | 10 am | 11 am | 12 pm | 1 pm | 2 pm | 3 pm | 4 pm | Close |
|---|---|---|---|---|---|---|---|---|---|
| Tail parameter $\alpha$ | 3.06 (0.65) | 3.25 (0.69) | 2.68 (0.57) | 3.30 (0.70) | 3.62 (0.77) | 3.51 (0.75) | 6.34 (1.35) | 3.03 (0.65) | 3.11 (0.66) |
| H0: $\alpha>2$ | 1.63 (0.45) | 1.81 (0.46) | 1.18 (0.38) | 1.85 (0.47) | 2.10 (0.48) | 2.02 (0.48) | 3.21 (0.50) | 1.60 (0.45) | 1.68 (0.45) |
| H0: $\alpha>4$ | -1.43 (0.00) | -1.08 (0.00) | -2.32 (0.00) | -0.99 (0.00) | -0.49 (0.00) | -0.65 (0.00) | 1.73 (0.46) | -1.50 (0.00) | -1.33 (0.00) |

**Panel B. Daily future price changes - Right tail**

|  | Open | 10 am | 11 am | 12 pm | 1 pm | 2 pm | 3 pm | 4 pm | Close |
|---|---|---|---|---|---|---|---|---|---|
| Tail parameter $\alpha$ | 2.58 (0.55) | 3.63 (0.77) | 4.34 (0.93) | 3.77 (0.80) | 4.20 (0.90) | 3.48 (0.74) | 4.96 (1.06) | 4.08 (0.87) | 3.64 (0.78) |
| H0: $\alpha>2$ | 1.05 (0.35) | 2.11 (0.48) | 2.53 (0.49) | 2.20 (0.49) | 2.46 (0.49) | 2.00 (0.48) | 2.80 (0.50) | 2.39 (0.49) | 2.11 (0.49) |
| H0: $\alpha>4$ | -2.59 (0.00) | -0.48 (0.00) | 0.37 (0.14) | -0.29 (0.00) | 0.22 (0.09) | -0.70 (0.00) | 0.91 (0.32) | 0.09 (0.04) | -0.47 (0.00) |

**Panel C. High-frequency future price changes - Left and right tails**

|  | Left tail | | | Right tail | | |
|---|---|---|---|---|---|---|
|  | 5-minute | 1-hour | 1-day | 5-minute | 1-hour | 1-day |
| Tail parameter $\alpha$ | 2.77 (0.01) | 2.83 (0.04) | 3.11 (0.66) | 2.42 (0.01) | 2.71 (0.04) | 3.64 (0.78) |
| H0: $\alpha>2$ | 7.94 (0.50) | 3.98 (0.50) | 1.68 (0.45) | 4.95 (0.50) | 3.55 (0.50) | 2.11 (0.49) |
| H0: $\alpha>4$ | -12.68 (0.00) | -5.61 (0.00) | -1.33 (0.00) | -18.64 (0.00) | -6.46 (0.00) | -0.47 (0.00) |

*Note :* this table gives the tail parameter estimates for the left tail (Panel A) and the right tail (Panel B) of the distribution of daily price changes and for the left and right tails (Panel C) of the distribution of 5-minute, 1-hour and daily price changes. It also provides a test of the existence of the moments of the distribution. The first line of the table gives the tail parameter estimate obtained with the method developed by Huisman *et al* (2001) with the standard error below in parentheses. The second and third lines give the results of a test of the existence of the second moment (the variance) and the fourth moment (the kurtosis) with the *p*-value below in parentheses. As the tail parameter corresponds to the highest moment defined for the distribution, the null hypotheses are defined as follows: $H_0: \alpha > 2$ and $H_0: \alpha > 4$. To define the daily price change, the starting time (which is equal to the ending time on the

following day) varies from 9 am (opening of the market) to 5 pm (closing of the market). Data are price changes of the FTSE 100 futures contract over the year 2000.

**Table 4. Margin levels based on daily price changes for the FTSE 100 futures contract.**

**Panel A. Long position**

| Probability (waiting period) | Model | Open | 10 am | 11 am | 12 pm | 1 pm | 2 pm | 3 pm | 4 pm | Close |
|---|---|---|---|---|---|---|---|---|---|---|
| 95% (20 days) | Gaussian | 2.21 | 2.06 | 2.00 | 2.05 | 1.97 | 2.15 | 2.05 | 1.94 | 2.17 |
| | Extreme value | 1.85 | 1.95 | 1.89 | 1.84 | 2.04 | 1.83 | 1.85 | 1.95 | 2.05 |
| | Historical | 1.90 | 1.87 | 2.23 | 2.08 | 2.34 | 2.14 | 2.04 | 2.28 | 2.28 |
| | GARCH | 2.04 | 1.95 | 2.16 | 2.03 | 2.25 | 2.10 | 1.96 | 2.34 | 2.24 |
| 99% (100 days) | Gaussian | 3.11 | 2.90 | 2.82 | 2.89 | 2.78 | 3.03 | 2.88 | 2.73 | 3.05 |
| | Extreme value | 2.94 | 3.22 | 3.12 | 2.70 | 2.78 | 2.42 | 2.26 | 2.74 | 2.93 |
| | Historical | 2.98 | 3.23 | 3.06 | 2.76 | 2.90 | 2.89 | 2.51 | 3.19 | 3.25 |
| | GARCH | 3.12 | 3.15 | 2.85 | 2.89 | 2.93 | 2.92 | 2.67 | 3.13 | 3.27 |
| 99.60% (250 days) | Gaussian | 3.54 | 3.30 | 3.21 | 3.29 | 3.16 | 3.45 | 3.28 | 3.11 | 3.48 |
| | Extreme value | 3.83 | 4.29 | 4.15 | 3.35 | 3.32 | 2.84 | 2.54 | 3.32 | 3.59 |
| | Historical | 3.59 | 3.39 | 3.41 | 3.01 | 3.01 | 3.10 | 2.71 | 3.31 | 3.45 |
| | GARCH | 3.23 | 4.25 | 4.16 | 3.38 | 3.02 | 3.16 | 2.92 | 3.52 | 4.10 |
| 99.80% (500 days) | Gaussian | 3.84 | 3.58 | 3.48 | 3.57 | 3.43 | 3.74 | 3.55 | 3.37 | 3.77 |
| | Extreme value | 4.67 | 5.32 | 5.15 | 3.95 | 3.79 | 3.20 | 2.77 | 3.84 | 4.18 |
| | Historical | na | na | na | na | na | na | na | na | na |
| | GARCH | 4.03 | 4.46 | 4.81 | 3.82 | 3.44 | 3.28 | 3.14 | 3.74 | 4.14 |

**Panel B. Short position**

| Probability (waiting period) | Model | Open | 10 am | 11 am | 12 pm | 1 pm | 2 pm | 3 pm | 4 pm | Close |
|---|---|---|---|---|---|---|---|---|---|---|
| 95% (20 days) | Gaussian | 2.13 | 1.98 | 1.94 | 1.99 | 1.91 | 2.09 | 1.97 | 1.88 | 2.11 |
| | Extreme value | 1.70 | 1.76 | 1.80 | 1.65 | 1.96 | 1.74 | 1.77 | 2.06 | 1.94 |
| | Historical | 1.85 | 1.72 | 1.75 | 1.73 | 2.06 | 2.03 | 1.92 | 2.19 | 2.10 |
| | GARCH | 1.66 | 1.63 | 1.76 | 1.72 | 1.94 | 2.06 | 1.89 | 2.21 | 2.18 |
| 99% (100 days) | Gaussian | 3.03 | 2.82 | 2.76 | 2.83 | 2.72 | 2.97 | 2.80 | 2.67 | 2.99 |
| | Extreme value | 2.69 | 2.91 | 2.98 | 2.41 | 2.67 | 2.31 | 2.16 | 2.89 | 2.77 |
| | Historical | 2.76 | 2.82 | 2.67 | 2.47 | 2.82 | 2.50 | 2.37 | 2.78 | 2.77 |
| | GARCH | 3.08 | 2.97 | 2.65 | 2.40 | 2.61 | 2.49 | 2.30 | 2.79 | 2.86 |
| 99.60% (250 days) | Gaussian | 3.42 | 3.46 | 3.22 | 3.15 | 3.23 | 3.10 | 3.39 | 3.20 | 3.05 |
| | Extreme value | 3.87 | 3.87 | 3.97 | 2.99 | 3.18 | 2.71 | 2.42 | 3.51 | 3.40 |
| | Historical | 3.70 | 3.01 | 2.90 | 2.58 | 2.97 | 2.70 | 2.48 | 2.96 | 3.20 |
| | GARCH | 3.51 | 3.10 | 2.82 | 2.64 | 3.38 | 2.80 | 2.33 | 2.92 | 2.89 |
| 99.80% (500 days) | Gaussian | 3.76 | 3.50 | 3.42 | 3.51 | 3.37 | 3.68 | 3.47 | 3.31 | 3.71 |
| | Extreme value | 4.80 | 4.80 | 4.93 | 3.53 | 3.63 | 3.05 | 2.63 | 4.06 | 3.96 |
| | Historical | na | na | na | na | na | na | na | na | na |
| | GARCH | 3.73 | 3.20 | 3.35 | 2.80 | 3.79 | 3.01 | 2.53 | 3.06 | 3.00 |

*Note :* this table gives the margin level for a long position (Panel A) and a short position (Panel B) for different probability levels ranging from 95% to 99.8% or equivalently different waiting periods ranging from 20 trading days (1 month) to 500 trading days (2 years). Different statistical models are used: three unconditional distributions (the Gaussian distribution, the extreme value distribution and the historical distribution) and a GARCH process. The historical estimates are not available (na) for out of sample inferences due to data unavailability. To define the daily price change, the starting time (which is equal to the ending time on the following day) varies from 9 am (opening of the market) to 5 pm (closing of the market). Data are price changes of the FTSE 100 futures contract over the year 2000.

**Table 5: Daily margin levels based on 5-minute, 1-hour and 1-day price changes for the FTSE 100 futures contract.**

**Panel A. Long position**

| Probability (waiting period) | Model | Frequency of price changes | | |
|---|---|---|---|---|
| | | 5 minutes | 1 hour | 1 day |
| 95% (20 days) | Gaussian | 3.07 | 2.52 | 2.07 |
| | Extreme value | 2.17 | 1.64 | 1.92 |
| | Historical | na | na | 1.75 |
| | GARCH | 1.63 | 1.29 | 2.12 |
| 99% (100 days) | Gaussian | 4.34 | 3.52 | 2.91 |
| | Extreme value | 3.18 | 3.06 | 2.79 |
| | Historical | na | na | 2.67 |
| | GARCH | 3.32 | 2.65 | 2.99 |
| 99.60% (250 days) | Gaussian | 4.95 | 4.00 | 3.31 |
| | Extreme value | 4.48 | 4.12 | 3.47 |
| | Historical | na | na | 2.90 |
| | GARCH | 4.36 | 3.44 | 3.53 |
| 99.80% (500 days) | Gaussian | 5.37 | 4.33 | 3.59 |
| | Extreme value | 5.81 | 5.08 | 4.10 |
| | Historical | na | na | na |
| | GARCH | 5.37 | 4.33 | 3.87 |

**Panel B. Short position**

| Probability (waiting period) | Model | Frequency of price changes | | |
|---|---|---|---|---|
| | | 5 minutes | 1 hour | 1 day |
| 95% (20 days) | Gaussian | 3.07 | 2.32 | 2.00 |
| | Extreme value | 2.33 | 1.47 | 1.82 |
| | Historical | na | na | 1.93 |
| | GARCH | 1.58 | 1.20 | 1.89 |
| 99% (100 days) | Gaussian | 4.34 | 3.32 | 2.84 |
| | Extreme value | 3.40 | 3.16 | 2.64 |
| | Historical | na | na | 2.66 |
| | GARCH | 3.22 | 2.25 | 2.68 |
| 99.60% (250 days) | Gaussian | 4.95 | 3.80 | 3.25 |
| | Extreme value | 4.57 | 4.31 | 3.32 |
| | Historical | na | na | 2.94 |
| | GARCH | 4.30 | 2.96 | 2.93 |
| 99.80% (500 days) | Gaussian | 5.37 | 4.13 | 3.53 |
| | Extreme value | 6.46 | 5.20 | 3.93 |
| | Historical | na | na | na |
| | GARCH | 5.16 | 3.20 | 3.16 |

*Note :* this table gives the daily margin levels for a long position (Panel A) and a short position (Panel B) for different probability levels ranging from 95% to 99.80% or equivalently different waiting periods ranging from 20 trading days (1 month) to 500 trading days (2 years). Three different frequencies are used to compute the price changes: 5 minutes, 1 hour and 1 day. Different statistical models are used: three unconditional distributions (the Gaussian distribution, the extreme value distribution and the historical distribution) and a GARCH process. The historical estimates are not available (na) due to the lack of a scaling formula or to data unavailability for out-of-sample inferences. Margin levels obtained with 5-minute price changes and 1-hour price changes are scaled to obtain daily margin levels. Margin levels obtained from daily price changes correspond to the average over the margin levels obtained with different time-intervals. Data are price changes of the FTSE 100 futures contract over the year 2000.